\documentstyle[sprocl,psfig]{article}

\catcode`\@=11
\def\@cite#1#2{{\if@cghi$\!^{#1}$\else$[{#1}]$\fi\if@tempswa\typeout
        {IJCGA warning: optional citation argument
        ignored: `#2'} \fi}}
\catcode`\@=\active

\def\pl#1#2#3{Phys.~Lett.~{\bf {#1}B} (19{#2}) #3}

\def\np#1#2#3{Nucl.~Phys.~{\bf B{#1}} (19{#2}) #3}

\def\cites#1#2{\cite{#1}$^-\,$\cite{#2}}

\def\wt{\widetilde}

\def\sst{\scriptscriptstyle}

\def\alg{$shs^E(8|4)$}
\def\mi#1#2#3{#1_{#2}\cdots #1_{#3}}

\def\nab{\nabla}
\def\del{\partial}

\newcommand{\w}[1]{\\[0.#1cm]}

\def\be{\begin{equation}}
\def\ee{\end{equation}\\[-.75cm] }
\def\ba{\begin{array}}
\def\ea{\end{array}}
\def\bea{\begin{eqnarray}}
\def\eea{\end{eqnarray}\\[-.75cm] }
\def\bd{\begin{document}}
\def\ed{\end{document}}

\let\bm=\bibitem

\let\la=\label
\def\nn{\nonumber}

\def\qq{\quad\quad}
\def\se{\;\;=\;\;}  \def\de{\;\;:=\;\;}

\def\ft#1#2{{\textstyle{{\scriptstyle #1}\over {\scriptstyle #2}}}}
\def\fft#1#2{{#1 \over #2}}

\def\sst#1{{\scriptscriptstyle #1}}
\def\smpl{{\tiny +}}
\def\smm{{\tiny -}}
\def\oneone{\rlap 1\mkern4mu{\rm l}}

\newcommand{\eq}[1]{Eq.~(\ref{#1})}
\newcommand{\sect}[1]{Sec.~(\ref{#1})}

\newcommand{\eqs}[2]{Eqs.~(\ref{#1})-(\ref{#2})}

\def\Hat#1{\widehat{#1}}

\def\a{\alpha}
\def\ad{\dot \alpha}
\def\ua{\underline{\alpha}}
\def\ub{\underline{\phantom{\alpha}}\!\!\!\beta}
\def\uc{\underline{\phantom{\alpha}}\!\!\!\gamma}
\def\una{\underline{a}}
\def\b{\beta}
\def\bd{\dot \beta}
\def\c{\gamma}
\def\C{\Gamma}
\def\cd{\dot\gamma}
\def\d{\delta}
\def\D{\Delta}
\def\dd{\dot\delta}
\def\e{\epsilon}
\def\vare{\varepsilon}
\def\he{\hat{\e}}
\def\f{\phi}
\def\tif{\varphi}
\def\F{\Phi}
\def\vf{\varphi}
\def\hvf{\hat{\vf}}
\def\y{\eta}
\def\i{\iota}
\def\k{\kappa}
\def\kb{\bar{\k}}
\def\l{\lambda}
\def\L{\Lambda}
\def\m{\mu}
\def\n{\nu}
\def\p{\pi}
\def\P{\Pi}
\def\bp{\bar{\pi}}
\def\r{\rho}
\def\s{\sigma}
\def\S{\Sigma}
\def\sb{\bar{\s}}
\def\t{\tau}
\def\th{\theta}
 \def\Th{\Theta}
 \def\vth{\vartheta}
 \def\tb{\bar\theta}
\def\x{\xi}
\def\xb{\bar{\x}}
\def\X{\Xeta}
\def\o{\omega}
\def\O{\Omega}
\def\bo{\bar{\o}}
\def\cV{{\cal V}}
\def\bcV{\bar{{\cal V}}}

\def\yb{\bar{y}}
\def\zb{\bar{z}}
\def\hD{\hat{D}}
\def\hd{\hat{d}}
\def\he{\hat{\varepsilon}}
\def\ho{\hat{\o}}
\def\hR{\hat{{\cal R}}}
\def\hf{\hat{\f}}
\def\thf{\hat{\vf}}
\def\hx{\hat{\x}}
\def\cR{{\cal R}}
\def\cF{{\cal F}}

\def\ns{\normalsize}

\def\ket#1{\left|#1\right>}

\def\sc#1#2{[\,#1\,,\,#2\,]_{\star}}

\def\alg{$shs^E(8|4)$}
\def\algN{$shs^E(N|4)$}
\def\aext{$\widehat{\cal A}$}
\def\bext{$\widehat{\cal B}$}
\def\alge{$\widehat{shs}{}^E(8|4)$}

\def\g{\u g}
\def\cM{{\cal M}}

\def\tc{
\begin{table}[t]
\begin{center}
{\footnotesize
\tabcolsep=1mm
\begin{tabular}{|c|cc|cccccccccccc|}\hline
& & & & & & & & & & & & & & \\
{\large${}_{\ell}\backslash s$} & $0$ & \ns{$\ft12$} & $1$
& \ns{$\ft32$} & $2$ &
\ns{$\ft52$} &
$3$ & \ns{$\ft72$} & $4$ & $\ft92$ & $5$ & $\ft{11}2$ & $6$ & $\cdots$ \\
& & & & & & & & & & & & & & \\ \hline
& & & & & & & & & & & & & & \\
$0$ & $35_{\smpl}\!+\!35_{\smm}$ & $56$ & $28$ & $8$ & $1$ &
& & & & & & & & \\
$1$ & $1\!+\!1$ & $8$ & $28$ & $56$ & $35_{\smpl}\!+\!35_{\smm}$
& $56$ & $28$ & $8$ & $1$ &
& & & &  \\
$2$ & & & & & $1$ & $8$ & $28$ & $56$ & $35_{\smpl}\!+\!35_{\smm}$
& $56$ & $28$ & $8$ & $1$  &  \\
$3$ & & & & & & & & & $1$ & $8$ & $28$ & $56$ &  $35_{\smpl}\!+\!35_{\smm}$
& $\cdots$ \\
$4$ & & & & & & & & & & & & & $1$ & $\cdots$ \\
$\vdots$ & & & & & & & & & & & & & &  \\ \hline
\end{tabular}}
\end{center}
\caption{{\small The $SO(3,2)\times SO(8)$ content of the symmetric
tensor product of two $N=8$ supersingletons. Each entry refers to the
$SO(8)$ content and the energy $E_0=s+1$, where $s$ is the spin.
The representations have been arranged into a tower of $OSp(8|4)$
supermultiplets labeled by a level number $\ell$ explained in \sect{sec:alg}.
The vertical bar separates the spin $s\geq 1$ physical modes of the
\alg\ gauge field $\o$ from the physical modes of the Weyl $0$-form
$\f$ introduced in \sect{sec:ga}
}}
\la{table}
\end{table}
}

\def\bw#1#2{\put(#1,#2){\circle{3}}}
\def\bdot#1#2{\put(#1,#2){{\tiny$\bullet$}}}
\def\fw#1#2{\put(#1,#2){$\diamondsuit$}}
\def\lc#1#2{\put(#1,#2){{\Large $\star$}}}
\def\aux#1#2{\put(#1,#2){{\large $\times$}}}
\def\cdo#1#2{\put(#1,#2){{\huge $......$}}}
\def\vdo#1#2{\put(#1,#2){{\Huge $\vdots$}}}
\def\va#1#2{\put(#1,#2){\vector(0,-1){4}}}
\def\ha#1#2{\put(#1,#2){\vector(-1,0){4}}}

\def\tw{
 \begin{figure}[!t]
 \begin{center}
 \unitlength=.6mm
\begin{picture}(120,120)(0,-10) \put(0,0){\circle*{3}}
\bw{10}{10}\bdot{9}{9.2}\bw{20}{20}\bw{30}{30}\bdot{29}{29.2}
\bw{40}{40}\bw{50}{50}\bdot{49}{49.2}\bw{60}{60}\bw{70}{70}
\bdot{69}{69.3}\bw{80}{80}
\fw{-2}{8.5}\fw{8}{18.5}\fw{18}{28.5}\fw{28}{38.5}\fw{38}{48.5}
\fw{48}{58.5}\fw{58}{68.5}\fw{68}{78.5}
\fw{8}{-1.5}\fw{18}{8.5}\fw{28}{18.5}\fw{38}{28.5}\fw{48}{38.5}
\fw{58}{48.5}\fw{68}{58.5}\fw{78}{68.5}
\lc{18}{-2}\lc{28}{8}\lc{38}{18}\lc{48}{28}\lc{58}{38}
\lc{68}{48}\lc{78}{58}
\lc{-2}{18.5}\lc{8}{28.5}\lc{18}{38.5}\lc{28}{48.5}\lc{38}{58.5}
\lc{48}{68.5}\lc{58}{78.5}
\aux{-2.5}{28.5}\aux{-2.5}{38.5}\aux{-2.5}{48.5}\aux{-2.5}{58.5}
\aux{-2.5}{68.5}\aux{-2.5}{78.5}
\aux{7.5}{38.5}\aux{7.5}{48.5}\aux{7.5}{58.5}\aux{7.5}{68.5}\aux{7.5}{78.5}
\aux{17.5}{48.5}\aux{17.5}{58.5}\aux{17.5}{68.5}\aux{17.5}{78.5}
\aux{27.5}{58.5}\aux{27.5}{68.5}\aux{27.5}{78.5}
\aux{37.5}{68.5}\aux{37.5}{78.5} \aux{47.5}{78.5}
\aux{27.5}{-1.5}\aux{37.5}{-1.5}\aux{47.5}{-1.5}\aux{57.5}{-1.5}
\aux{67.5}{-1.5}\aux{77.5}{-1.5}
\aux{37.5}{8.5}\aux{47.5}{8.5}\aux{57.5}{8.5}\aux{67.5}{8.5}\aux{77.5}{8.5}
\aux{47.5}{18.5}\aux{57.5}{18.5}\aux{67.5}{18.5}\aux{77.5}{18.5}
\aux{57.5}{28.5}\aux{67.5}{28.5}\aux{77.5}{28.5}
\aux{67.5}{38.5}\aux{77.5}{38.5} \aux{77.5}{48.5}
\vdo{-2}{90}\vdo{-2}{94.4} \vdo{8}{90}\vdo{8}{94.4}
\vdo{18}{90}\vdo{18}{94.4} \vdo{28}{90}\vdo{28}{94.4}
\vdo{38}{90}\vdo{38}{94.4} \vdo{48}{90}\vdo{48}{94.4}
\vdo{58}{90}\vdo{58}{94.4} \vdo{68}{90}\vdo{68}{94.4}
\vdo{78}{90}\vdo{78}{94.4}
\cdo{90}{80}\cdo{90}{70}\cdo{90}{60}\cdo{90}{50}\cdo{90}{40}
\cdo{90}{30}\cdo{90}{20}\cdo{90}{10}\cdo{90}{0} \put(-7,110){$n$}
\put(115,-8){$m$} \put(0,103){\vector(0,1){20}}
\put(110,0){\vector(1,0){20}}
\va{50}{47.5}\ha{47.5}{50}\va{40}{47}\ha{47.5}{40}
\va{40}{37.5}\ha{37.5}{40}\va{30}{37}\ha{37.5}{30}
\va{30}{27.5}\ha{27.5}{30}\va{20}{27}\ha{27.5}{20}
\va{20}{17.5}\ha{17.5}{20}\va{10}{17}\ha{17.5}{10}
\end{picture}
\end{center} \caption{{\small Each entry of the integer grid,
$m,n=0,1,2,...$, represents a component $\o(m,n;\th)$ of the \alg\
valued connection $1$-form $\o$~: the $\bullet$ denote the spin $s=1$
gauge fields; a {\Large $\circ$} or a {\Large
$\circ$}$\hspace{-5.2pt}^{_{\bullet}}$ denotes a generalized vierbein;
a $\diamondsuit$ denotes a generalized gravitino; a $\star$ denotes a
generalized Lorentz connection and the $\times$'s denote the remaining
auxiliary gauge fields. The $SO(8)$ content is suppressed in this Figure,
but it can be read off from Table \ref{table}. The auxiliary gauge fields
along the northwest-southeast diagonal denoted $\times$'s and $\star$'s
are given in terms of the dynamical gauge fields in the middle of
that diagonal denoted by {\large $\circ$}'s and the $\diamondsuit$'s.
The arrows indicate the action of supersymmetry on the members of
the level $\ell=2$ multiplet. }}
\la{wfig}
 \end{figure} }

\def\wt#1#2{\lc{#1}{#2}}
\def\fa#1#2{\aux{#1}{#2}}
\def\eom#1#2{\put(#1,#2){\circle*{3}}}
\def\ff#1#2{\fw{#1}{#2}}

\def\tf{
 \begin{figure}[!h]
 \begin{center}
 \unitlength=.6mm
\begin{picture}(120,120)(0,-10)
\put(-7,110){$n$}\put(115,-8){$m$}
\put(0,103){\vector(0,1){20}}\put(110,0){\vector(1,0){20}}
\wt{-2}{18.5} \wt{-2}{28.5} \wt{-2}{38.5} \wt{-2}{48.5} \wt{-2}{58.5}
\wt{-2}{68.5} \wt{-2}{78.5}
\wt{18}{-1.5} \wt{28}{-1.5} \wt{38}{-1.5} \wt{48}{-1.5} \wt{58}{-1.5}
\wt{68}{-1.5} \wt{78}{-1.5}
\fa{7.5}{8.5} \fa{7.5}{18.5} \fa{7.5}{28.5} \fa{7.5}{38.5} \fa{7.5}{48.5}
\fa{7.5}{58.5} \fa{7.5}{68.5} \fa{7.5}{78.5}
\fa{17.5}{8.5}\fa{17.5}{18.5} \fa{17.5}{28.5} \fa{17.5}{38.5}
\fa{17.5}{48.5} \fa{17.5}{58.5} \fa{17.5}{68.5} \fa{17.5}{78.5}
\fa{27.5}{8.5}  \fa{27.5}{18.5} \fa{27.5}{28.5} \fa{27.5}{38.5}
\fa{27.5}{48.5} \fa{27.5}{58.5} \fa{27.5}{68.5} \fa{27.5}{78.5}
\fa{37.5}{8.5}  \fa{37.5}{18.5} \fa{37.5}{28.5} \fa{37.5}{38.5}
\fa{37.5}{48.5} \fa{37.5}{58.5} \fa{37.5}{68.5} \fa{37.5}{78.5}
\fa{47.5}{8.5}  \fa{47.5}{18.5} \fa{47.5}{28.5} \fa{47.5}{38.5}
\fa{47.5}{48.5} \fa{47.5}{58.5} \fa{47.5}{68.5} \fa{47.5}{78.5}
\fa{57.5}{8.5}  \fa{57.5}{18.5} \fa{57.5}{28.5} \fa{57.5}{38.5}
\fa{57.5}{48.5} \fa{57.5}{58.5} \fa{57.5}{68.5} \fa{57.5}{78.5}
\fa{67.5}{8.5}  \fa{67.5}{18.5} \fa{67.5}{28.5} \fa{67.5}{38.5}
\fa{67.5}{48.5} \fa{67.5}{58.5} \fa{67.5}{68.5} \fa{67.5}{78.5}
\fa{77.5}{8.5}  \fa{77.5}{18.5} \fa{77.5}{28.5} \fa{77.5}{38.5}
\fa{77.5}{48.5} \fa{77.5}{58.5} \fa{77.5}{68.5} \fa{77.5}{78.5}
\ff{-2}{8.5}\ff{8}{-1.5}
\bw{0}{0}
\cdo{90}{80}\cdo{90}{70}\cdo{90}{60}\cdo{90}{50}\cdo{90}{40}
\cdo{90}{30}\cdo{90}{20}\cdo{90}{10}\cdo{90}{0}
\vdo{-2}{90}\vdo{-2}{94.4} \vdo{8}{90}\vdo{8}{94.4}
\vdo{18}{90}\vdo{18}{94.4} \vdo{28}{90}\vdo{28}{94.4}
\vdo{38}{90}\vdo{38}{94.4} \vdo{48}{90}\vdo{48}{94.4}
\vdo{58}{90}\vdo{58}{94.4} \vdo{68}{90}\vdo{68}{94.4}
\vdo{78}{90}\vdo{78}{94.4}
\end{picture}
\end{center}
\caption{{\small Each entry of the integer grid, $m,n=0,1,2,...$,
 represents a component $\f(m,n;\th)$ of $\f$~: the {\Large $\circ$}
 denotes the scalars, the two $\diamondsuit$'s
 denote the spin $\ft12$ fields; a $\star$ denotes a  chiral, auxiliary
 component $\f(2s,0,\th)$ ($s=1,\ft32,...$) which is equal to the spin $s$
 Weyl tensor given in \eq{weyl1}; the $\times$'s denote  auxiliary
 components which are given in terms of derivatives of the chiral
 components according to \eq{fsolv}. }}
\la{fifig}
\end{figure}
}

\begin{document}

\quad

\vspace{-1in}

\hfill{CTP TAMU-7/99}

\hfill{hep-th/9903020}

\vspace{1in}

\title{ HIGHER SPIN ${\bf N=8}$ SUPERGRAVITY IN ${\bf AdS_4}$
\footnote{Contribution to
 the John Hopkins Workshop, August 1998, Chalmers, Sweden.}
}

\author{ E. SEZGIN }

\address{Center for Theoretical Physics, Texas A\&M University,
College Station, \\ TX 77843, USA}

\author{P. SUNDELL}

\address{Department of Engineering Sciences, Physics and Mathematics
Karlstad University, S-651 88 Karlstad, Sweden}

\maketitle \abstracts {We review the basic structure of the higher spin
extension of $D=4,N=8$ AdS supergravity. The theory is obtained by
gauging the higher spin superalgebra $shs^E(8|4)$ by a procedure
pioneered by Vasiliev. The algebra $shs^E(8|4)$ is a subalgebra of the
enveloping algebra of $OSp(8|4)$. The physical states of the theory are
in one to one correspondence with the symmetric product of two
$OSp(8|4)$ singletons. This singleton theory, which may be viewed in a
certain limit as the supermembrane theory on $AdS_4
\times S^7$, is expected to describe the dynamics of the higher spin
theory. Thus, the higher spin $N=8$ supergravity on $AdS_4$ is
conjectured to describe the field theory limit of $M$-theory on
$AdS_4\times S^7$.}

\pagebreak


\section{Introduction}


A consistent interacting field theory of fields with spin higher than
two has long interested the high energy physicists. Many attempts to
construct such theories encountered severe difficulties in the past and
led to no-go theorems. However, no-go theorems are usually based on a
number of assumptions and such theorems may as well disappear when some
of those assumptions are relaxed. Indeed, this is just what happened in
the search for higher spin field theories as well. Firstly, and perhaps
not surprisingly, the number of higher spin fields participating in the
full theory need to be infinite, as opposed to one higher spin field at
a time considered previously, and secondly, and that is the surprising
part, the spacetime in which the theory is formulated must admit AdS
space as a solution \cite{fv0,fv1} rather than Minkowskian space
assumed in previous attempts.

The occurrence of infinitely many fields is very suggestive of an
underlying extended object theory. String theory in $AdS_4$ has been
considered briefly in \citelow{fl} but the results appear to be
inconclusive. In a separate line of development, the eleven dimensional
supermembrane theory appeared in 1987 \cite{bst1}. Soon after, it was
suggested \cite{duff1} that the $AdS_4$ supersingleton may play a role
in its description. This is the ultra short representation of the
$AdS_4$ supergroup $OSp(4|8)$, consisting of 8 bosonic and 8 fermionic
states \cite{g1,ns} which cannot be described in terms of local fields
in the bulk of $AdS_4$. Soon after that, a connection between the
$OSp(8|4)$ supersingleton field theory, assumed to arise from the
$AdS_4\times S^7$ compactification of the supermembrane, and a higher
spin supergravity theory in $AdS_4$ bulk was conjectured in
\citelow{bst0,bsst1,bsst2}. This conjecture was especially aided by the
remarkable property of the singletons, namely the fact that the
symmetric product of two supersingletons yields an infinite tower of
massless higher spin states \cite{ff1}. The form of this result for
$OSp(8|4)$ was spelled out in \citelow{bsst2}, and it was argued that
all of these states should arise in the quantum supermembrane theory
\cite{bst0,bsst1,bsst2}.

These developments provided further motivation for the analysis of the
supermembrane on various manifolds of the form $AdS_4 \times M_7$,
where $M_7$ is a positive curvature Einstein manifold, and various
issues arising in the expansion of the supermembrane action around
these vacua \cite{mc1,mc2,dps}. These were the early developments on
the subject of bulk AdS and boundary CFT connection in the context of
the $D=11$ supermembrane its simplest form. The AdS/CFT connection has
acquired a deeper and wider significance in the light the developments
of the last year which were triggered by the work of Maldacena
\cite{jm}.

Turning to the story of higher spin fields, interestingly enough, at
the same period when as the supermembrane/singletons/higher-spin-fields
connections were being considered, Fradkin and Vasiliev \cite{fv0,fv1}
were in the course of developing a higher spin gauge theory in its own
right (see \citelow{fv0} for references to earlier work). These authors
succeeded in constructing interacting field theories for higher spin
fields. As mentioned above, the previous difficulties in constructing
higher spin theories were bypassed by formulating the theory in $AdS_4$
and by considering an infinite tower of gauge fields controlled by
various higher spin algebras based on certain infinite dimensional
extensions of $AdS_4$ superalgebras. An intriguing property of the
higher spin field theories, which again is very suggestive of an
underlying extended object, is the fact that the interactions are
non-local. In particular, the $AdS_4$ radius can not be taken to
infinity since its positive powers occur in the higher spin
interactions and therefore one can not take a naive Poincar\'e limit.

In a series of papers Vasiliev \cites{v1}{v10} pursued the program of
constructing the AdS higher spin gauge theory and simplified the
construction considerably. In \citelow{v4} the spin 0 and 1/2 fields
were introduced to the system at the level of the equations of motion
within the framework of free differential algebras. The equations of
motion were furthermore cast into an elegant geometrical form in
\citelow{v7,v10} by extending the higher spin algebra to include new
auxiliary spinorial variables. For $N=1$, $R\wedge R$ type actions have
been considered \cite{fv0,fv1}, but one drawback of these actions is
that the spin $s \leq 1$ sector of the theory does not fit in a natural
and geometrical way into the part of the action that describes the
fields with spin $s \geq 3/2$.

Interestingly enough, applying the formalism of Vasiliev to a suitable
higher spin algebra that contains the maximally extended super AdS
algebra $OSp(8|4)$, the resulting spectrum of gauge fields and spin
$s\leq \ft12$ fields coincide with the massless states resulting from
the symmetric product of two $OSp(8|4$ supersingletons \cite{bsst1}. In
a recent paper \cite{ss1}, we examined the Vasiliev theory of higher
spin fields (which is applicable to a wide class of higher spin
superalgebras) and determine the precise manner in which the $N=8$ de
Wit-Nicolai gauged supergravity \cite{dn1} which is the gauged version
of the Cremmer-Julia $N=8$ supergravity in four dimensions
\cite{cj1} can be described within this framework. We showed how
this embedding works in the higher spin AdS supergravity based on the
higher spin superalgebra known as \alg \cite{kv1,kv2}.

The most natural next step in this program is to compare the
interactions in the higher study with those of gauged $N=8$
supergravity, and to search for the $E_7/SU(8)$ structure \cite{cj1} in
the full theory. Even more tantalizing is the prospects of generating
the full content of the higher spin theory from the $OSp(8|4)$
singleton field theory which lives in the boundary of $AdS_4$.

The purpose of this report is far more modest, namely to give a concise
review of the work presented in \citelow{ss1}, and to emphasize the
elegant geometrical picture that underlies the higher spin field
equations. In Sec. 2, we discuss the higher spin algebra, its gauging
and the constraints which encode the dynamics. In Sec. 3, we discuss
the linearization of the higher spin equations around the $AdS_4$
vacuum, the spectrum of physical states, their equations of motion and
the comparison with the linearized field equations of the gauged $N=8,
D=4 $ supergravity. The uniqueness of the constraints is discussed in
Sec. 4. Further comments on the results and on open problems are
collected in Sec. 5.


\section{The Higher Spin Gauge Theory} \la{sec:hs}



\subsection{The Higher Spin Algebra} \la{sec:alg}


The higher spin algebra \alg\ arises most naturally as a truncation of
an extended higher algebra \alge\ . The latter is the algebra of
polynomials in the bosonic $SO(3,2)$ Majorana spinors $Y_{\ua}$ and
$Z_{\ua}$ ($\ua=1,...,4$) and the real, fermionic $SO(8)$ vector
$\th^i$ which obey the following associative but non-commutative
$\star$ product rule \cite{v10}

\be
P(Z,Y)\,\star\,Q(Z,Y)\se \int d^4U\; d^4V ~ P(Z+U,Y+U) \;
Q(Z-V,Y+V) e^{-iU^{\ua}V_{\ua}} \ ,
\la{star}
\ee

where the integral is normalized such that $1\star P=P$. This formula
in particular implies

\bea
Z_{\ua}\star Z_{\ub}&\se& Z_{\ua}Z_{\ub}-iC_{\ua\ub}\ ,\qq
Z_{\ua}\star Y_{\ub}\se Z_{\ua}Y_{\ub}+iC_{\ua\ub}\ ,
\nn\w2
Y_{\ua}\star Z_{\ub}&\se& Y_{\ua}Z_{\ub}-iC_{\ua\ub}\ ,\qq
Y_{\ua}\star Y_{\ub}\se Y_{\ua}Y_{\ub}+iC_{\ua\ub}\ ,
\eea

where $C_{\ua\ub}$ is the anti-symmetric charge conjugation matrix. The
product rule \eq{star} is isomorphic to the normal ordered product of a
set of harmonic oscillators formed out of the spinor variables. The
$\star$ product between $\th$'s is that of the $SO(8)$ Clifford
algebra, so that for example $\th^i\star
\th^j=\d^{ij}+\th^i\th^j$.

We next define the map

\be
\t P(Z;Y,\th)\de P(-iZ;iY,i\th)\ ,
\la{t}
\ee

which acts as a graded anti-homomorphism of the $\star$ algebra:

\be
\t (P\star Q)\se (-1)^{PQ} \t(Q)\star \t(P)\ .
\la{ah}
\ee

The extended higher spin algebra \alge\ is by definition the space of
Grassmann {\it even} polynomials $P(Z;Y,\th)$ obeying \cite{kv2}

\be
\widehat{shs}{}^E(8|4)\de \left\{P(Z;Y,\th)\left|\ \t P= -P\ ,\
P^{\dagger}= -P\right.\right\}\ ,
\la{shse}
\ee

with Lie bracket

\be
\sc{P}{Q}\de P\star Q- Q\star P\ .
\la{lb}
\ee

The higher spin algebra \alg\ is defined by \cite{kv2}

\be
shs^E(8|4) \de  \widehat{shs}{}^E(8|4)|_{Z=0}\ .
\la{shs}
\ee

The $OSp(8|4)$ subalgebra generators are

\be
OSP(8|4)\se \left\{\ Y_{\ua}Y_{\ub}\,,\ Y_{\ua}\th^i\,,\ \th^i\th^j\
\right\}\ .
\la{osp}
\ee

The algebra \alg\ splits under $OSp(8|4)$ into levels labeled by
$\ell=0,1,...$ and the $\ell$'th level generators are given by a
homogeneous polynomials of degree $4\ell+2$. The $0$'th level is
$OSp(8|4)$. Actually, \alg\ is the space of odd polynomials in
$OSp(8|4)$ generators which are fully symmetric in $SO(3,2)$ spinor
indices and fully anti-symmetric in $SO(8)$ vector indices.


\subsection{Gauging \alge}\la{sec:ga}


The gauging of \alge\ proceeds by introducing a master gauge field
$\ho$ which is an \alge\ valued $1$-form on the space $\cM$ which is
the product of spacetime with the non-commutative $Z$ space with
coordinate $z^M=(x^\m,Z^{\ua})$, where $x^\m$ are the spacetime
coordinates. Introducing the total exterior derivative

\be
\hd\de  dz^M\del_M\se dx^\m\del_\m+ dZ^{\ua}\del_{\ua}\ ,
\la{hd}
\ee

we define the \alge\ valued connection $1$-form
\footnote{
$\ho_\a$ is related to the field $S_\a$ introduced by Vasiliev in
\citelow{v7} by $S_\a=z_\a+2i\ho_\a$.}

\bea
\ho&\se& dz^M \ho_M(x,Z;Y,\th)\ ,
\nn\w2
\t\ho&\se&-\ho\ ,\qq \ho^{\dagger}\se -\ho\ .
\la{hw}
\eea

The \alge\ valued curvature $2$-form $\hR$ is

\be
\hR\se \hd \ho-\ho \star \ho\ ,
\la{hr}
\ee

which obeys the Bianchi identity

\be
\hD\hR\de \hd\hR-\sc{\ho}{\hR}\se0\ .
\la{hbi}
\ee

The \alge\  gauge transformations are

\be
\d_{\he}\ho\se \hD\he\ ,\qq
\d_{\he}\hR\se \sc{\he}{\hR}\ ,
\la{hgt}
\ee

where $\he$ is an \alge\ valued parameter.

The gauging of \alge\ differs from ordinary Yang-Mills theory in two
significant ways. Firstly, $\hR$ is a non-local $Z^{\ua}$ function of
$\ho$ due to the $\star$ operation in the quadratic term $\ho\star
\ho$. Imposing constraints on $\hR$, this non-locality in $Z$-space
manifests itself as non-locality in spacetime in the form of higher
derivative dependence on the fields. Secondly, viewing the connection
$\ho$ as a map from $\cM$ to the internal space coordinatized by
$(Y,\th)$, the values assumed by $\ho$ at a given point in $\cM$ depend
on the $Z$-coordinates of the point.

We define the \alg\ valued gauge field $\o$ and its curvature $\cR$ by

\bea
\o&\de&i^*\ho\se dx^\m \o_\m(x;Y,\th) \ ,
\la{w}\w2
\cR &\de&  i^*\hR\se\ft12 dx^\m\wedge dx^\n \cR_{\m\n}(x;Y,\th)\ ,
\la{r}
\eea

where $i: shs^E(8|4) \hookrightarrow \widehat{shs}{}^E(8|4)$ is the
embedding map. From the definition \eq{shse} of \alg\ it then follows
that $\o$ has the expansion

\bea
\o &\se& \fft{1}{2i}\sum_{k=0}^{\infty}\left(
\quad\;\sum_{m+n=4k}\;\; \left(\ft1{2!} \o_{ij}(m,n)\th^{ij}
+\ft1{6!}\o_{i_{1}\cdots i_{6}}(m,n)\th^{i_{1}\cdots i_{6}} \right)
\right. \nn\w2
&&+\sum_{m+n=4k+1}\left( \o_{i}(m,n)\th^{i}
+\ft1{5!}\o_{i_{1}\cdots i_{5}}(m,n)\th^{i_{1}\cdots i_{5}}\right)\nn\w2
&&+\sum_{m+n=4k+2}\left( \o(m,n)
+\ft1{4!}\o_{i_{1}\cdots i_{4}}(m,n)\th^{i_{1}\cdots i_{4}}
+\ft1{8!}\o_{i_{1}\cdots i_{8}}(m,n)\th^{i_{1}\cdots i_{8}}\right)\nn\w2
&&+\left. \sum_{m+n=4k+3}\left( \ft1{3!}\o_{ijk}(m,n)\th^{ijk}
+\ft1{7!}\o_{i_{1}\cdots i_{7}}(m,n)\th^{i_{1}\cdots i_{7}}\right)\right)
\ ,\la{wexp}
\eea
\w2
where we have used the notation

\be
P(m,n):=\ft1{m!n!}P_{\mi\a1m\mi{\ad}1n}y^{\a_1}
\cdots y^{\a_m}\yb^{\ad_1}\cdots
\yb^{\ad_n}\ .
\la{pmn}
\ee

Thus bosonic gauge fields are always in the $1$, $28$ and $35_++35_-$
representations of $SO(8)$, while the fermionic fields are always in
$8$ and $56$ representations.

To realize \alg\ invariant dynamics irreducibly on a set of fields in
spacetime, the \alge\ valued curvature $\hR$ must be constrained. There
exists a set of constraints which achieves this while maintaining
spacetime diffeomorphism invariance. The solution of these constraints
involve a $0$-form master field $\hf$ subject to certain conditions.
The field $\hf$ essentially consists of the non-vanishing components of
the spacetime \alg\ curvatures and the spin $s=0,\ft12$ fields listed
in Table \ref{table}. The required properties of $\hf$ turn out to be

\be
\t \hf\se \bp \hf\ ,\qq \hf^{\dagger}\se \pi(\hf)\star \C\ ,
\la{hfi}
\ee

where

\be
\C\se\th^1\th^2\cdots\th^8
\la{c}
\ee

is the $SO(8)$ chirality matrix, and $\pi$ and $\bp$ are homomorphisms
of the $\star$ algebra defined by

\bea
\pi P(z,\zb;y,\yb;\th)&\de& P(-z,\zb;-y,\yb;\th)
\ ,\nn\w2
\bp P(z,\zb;y,\yb;\th)&\de& P(z,-\zb;y,-\yb;\th)\ .
\la{pi}
\eea

In \eq{hfi} we have introduced an $SO(3,1)$ covariant notation where
$y_\a$ and $z_\a$ are complex, two-component Weyl spinors with
hermitian conjugates $\yb_{\ad}:=(y_\a)^{\dagger}$ and
$\zb_{\ad}:=(z_\a)^{\dagger}$ and $Y_{\ua}:=(y_{\a},\yb^{\ad})$ and
$Z_{\ua}:=(z_\a,-\zb^{\ad})$. This amounts to formulating the \alg\
invariant theory in an $SO(3,1)$ basis.

At $Z=0$ we find that the field

\be
\f(x;Y,\th)\de i^*\hf
\la{fi}
\ee

has the expansion

\be
\f \se \sum_{m>n\geq 0}
\left[\f(m,n;\th)+(-1)^{n}\f(m,n;\th)^{\dagger}\star \C
\right]+ \sum_{m\geq 0}\f(m,m;\th)\ ,
\la{fiexp}
\ee
\w2

where we have used the notation \eq{pmn} and where the complex fields
$\f(m,n;\th)$  have the $\th$-expansion

\be
\!\!\!\f(m,\!n;\th)=\left\{\ba{ll}
 \f(m,\!n)\! +\!\ft1{4!}\f_{i_{1}\cdot
i_{4}}(m,\!n)\th^{i_{1}\cdot i_{4}}\! +\!
\ft1{8!}\f_{i_{1}\cdot
i_{8}}(m,\!n)\th^{i_{1}\cdot i_{8}} &\quad{\footnotesize m\smm n=0\
\mbox{mod}\ 4}
\w3
\ft1{3!}\f_{ijk}(m,n)\th^{ijk} +\ft1{7!}\f_{i_{1}\cdots
i_{7}}(m,n)\th^{i_{1}\cdots i_{7}} &\quad{\footnotesize m\smm n=1\
\mbox{mod}\ 4}
\w3
\ft1{2!} \f_{ij}(m,n)\th^{ij} +\ft1{6!}\f_{i_{1}\cdots
i_{6}}(m,n)\th^{i_{1}\cdots i_{6}} &\quad{\footnotesize m\smm n=2\
\mbox{mod}\ 4}
\w3
\f_{i}(m,n)\th^{i}+\ft1{5!}\f_{i_{1}\cdots i_{5}}(m,n)\th^{i_{1}\cdots
i_{5}} &\quad{\footnotesize m\smm n=3\
\mbox{mod}\ 4}
\ea \right.
\la{fimn}
\ee

for $m>n$, and

\bea
\f(m,m;\th)&\se& \f(m,m)+\ft1{4!}\f_{ijkl}(m,m)\th^i\th^j\th^k\th^l +
(-1)^{m}\f(m,m)^{\dagger}\C
\ ,\nn\w2
\f^{ijkl}(m,m)^{\dagger}&\se& \ft{(-1)^m}{4!}\e_{ijklnpqr}
\f^{npqr}(m,m)\ .
\la{mm}
\eea

for $m=n$. Hence the bosonic fields are in the $1$, $28$ and
$35_++35_-$ representations of $SO(8)$ and the fermions are in the $8$
and $56$ representations. In particular we recognize that the scalar
fields $\f_{ijkl}(0,0)$ are the $35_{\smpl}\!+\!35_{\smm}$ real scalars
of the the $N=8$ supergravity multiplet \cite{cj1} (see Table
\ref{table}), and that these obey the $SU(8)$ invariant reality
condition which is crucial for the consistency of the $E_7/SU(8)$ coset
construction. We also find that $\f(0,0)$ is the complex scalar of the
level $1$ supermultiplet in Table \ref{table} and that $\f(1,0;\th)$
contains the fermions of level $0$ and $1$ in Table \ref{table}. As we
shall see in Sec. \ref{sec:vac}, the remaining spin $s\geq 1$
components of $\f$ are auxiliary fields that are given by the
derivatives of the spin $s\leq\ft12$ fields and certain non-vanishing
spacetime curvatures that are known as Weyl tensors.


\subsection{The Master Curvature Constraint}\la{sec:mcc}


A set of curvature constraints that realize the dynamics of the higher
spin gauge theory irreducibly are

\be
\hR_{\m\n}\se 0\ ,\qq  \hR_{\m\a}\se 0\ ,\qq \hR_{\a\bd}\se 0\ ,
\la{v}
\ee

and the 'twisted' reality condition

\be
\e^{\ad\bd}\hR_{\ad\bd}\se\e^{\a\b}\hR_{\a\b}\star K\ ,
\la{twr}
\ee

where the operator $K$ is defined by

\be
K\de \C\exp(iY^{\ua}Z_{\ua})\ .
\la{K}
\ee

The condition \eq{twr} implies that $\hR_{\a\b}$ can be expressed in
terms of the field $\hf$ introduced in \eq{hfi} as

\be
\hR_{\a\b}\se -\ft{i}2 \e_{\a\b}\hf\star \k\C\ ,
\la{rab}
\ee

where

\be
\k\de\exp(iy^\a z_\a)\ .
\la{k1}
\ee

This operator is an inner Kleinian operator\cite{v7} with the basic
property

\be
\k\star P(z,\zb;y,\yb;\th)\se \k\,P(y,\zb;z,\yb;\th)\ ,
\la{k2}
\ee

from which it follows that

\be
\k\star\k\se 1\ ,\qquad\k\star P\se \p( P)\star \k\ .
\la{k3}
\ee

The operator $K=\k\kb\C$ and from \eq{k3} it follows that

\be
K\star K\se 1\ ,\qquad K\star P \se \t^2 (P)\star K\ .
\la{K2}
\ee

We shall discuss the consequences of the twisted reality condition
\eq{twr} shortly. First, we need to establish the integrability and
\alge\ covariance of the constraints \eq{v} and \eq{rab}. To begin with
we note these constraints can be written concisely as

\be
\hR\se \ft{i}4(dz^\a\wedge dz_\a \hf\star \k\C +
d\zb^{\ad}\wedge d\zb_{\ad} \hf\star \kb)\ ,
\la{master}
\ee

Inserting the constraint \eq{master} into the Bianchi identity
\eq{hbi} one finds the following constraint on $\hf$

\be
\hD\hf\de \hd \hf -\ho\star \hf+\hf\star \bp \ho \se 0 \ .
\la{dhf}
\ee

Since $\hD^2\hf = 0$ there are no further conditions and the full set
of constraints given by \eq{master} and \eq{dhf} is integrable. The
constraint \eq{master} is invariant under the \alge\ transformation
\eq{hgt} provided that

\be
\d_{\he}\hf\se\he\star \hf-\hf\star \bp\he\ .
\la{hfgt}
\ee

These gauge transformations close, that is

\be
[\d_{\he_1},\d_{\he_2}]\hf\se -\d_{\sc{\he_1}{\he_2}}\hf\ .
\la{hc}
\ee

The gauge transformations \eq{hgt} and \eq{hfgt} also preserve
the constraint \eq{dhf}, since

\be
\d_{\he}\hD\hf\se \he\star \hD\hf-\hD\hf\star \bp\he\ .
\la{dd}
\ee

In summary, the full set of integrable constraints which define the
$N=8$ higher spin gauge theory based on \alg\ is:

\bea
\hD_{\a}\hf&\se&0\ ,
\la{e1}\w2
\hR_{\a\bd}&\se&0\ ,
\la{e2}\w2
\hR_{\a\b}&\se& -\ft{i}2\e_{\a\b}\hf\star \k\C\ ,
\la{e3}\w2
\hR_{\a\m}&\se&0\ ,
\la{e4}\w2
\hR_{\m\n}&\se&0\ ,
\la{e5}\w2
\hD_\m\hf&\se&0\ ,
\la{e6}\
\eea

We will discuss uniqueness in \sect{sec:dis}. While these equations
fully capture the higher spin theory based on
\alg, they contain many auxiliary fields. Their elimination and the
derivation of the full equations of motion in terms of a set of
dynamical fields is a highly non-trivial matter which has not been
solved yet. The structure of the equations become clear, however, by
linearizing them around a vacuum solution. One then finds that \eq{e1}
determines the $Z$ dependence of $\hf$ in terms of the initial
condition $\f$. Then \eqs{e2}{e3} determine the $Z$ dependence of
$\ho_{\a}$ in terms of $\f$ up to a pure gauge solution $\hD_{\a}\hx$
which can be used to impose the gauge

\be
\ho_\a|_{Z=0}\se 0 \ .
\la{hoa}
\ee

This amounts to that there are no spacetime degrees of freedom
associated with the initial conditions for $\ho_{\a}$. From \eq{e4} one
then solves for the $Z$ dependence of $\ho_{\m}$ in terms of $\f$ and
the initial condition $\o$. One is then left with
\eqs{e5}{e6}, which one can show are equivalent to

\bea
\cR &\de& i^*\hR\se\ft12 dx^\m\wedge dx^\n \cR_{\m\n}(x;Y,\th) \se 0\ ,
\nn\w2
D\f &\de& i^*\hD\hf\se dx^\m D_\m\f (x;Y,\th) \se 0\ ,
\la{z0}
\eea

where $\cR$ is the curvature two-form defined in \eq{r}. Observe that
\eq{z0} is the pull-back by $i^*$ of the full set of constraints
\eq{master} and \eq{dhf} to spacetime. If the $Z$ space had been an
ordinary commuting space, then this pull-back operation would have
implied the vanishing of the spacetime curvatures defined as

\bea
R&\de& d\o-\o\star\o\ ,
\nn\w2
d_{\o}\f&\de&d\f-\o\star\f+ \f\star\bp\o \ .
\la{rd}
\eea

The non-locality of the $Z$ space, however, implies that $\cR$ and
$D\f$ depend on all the coefficients of the Taylor expansions of
$\ho_\m$ and $\hf$ around $Z=0$, which in turn are determined in terms
of $\o$ and $\f$ as explained above. Hence $\cR$ and $D\f$ are infinite
expansions of the form $\cR=R+...$ and $D\f=d_\o\f+...$ in powers of
the $0$-form $\f$. Therefore \eq{z0} gives a set of non-trivial
constraints on the spacetime curvatures $R$ and $d_{\o}\f$.

The constraints \eq{z0} are invariant under the residual, gauge
transformations
\footnote{ \eq{hoa} implies that $\ho_\a\equiv z^\b\ho_{\a,\b}(x,Z;Y,\th)+
\zb^{\bd}\ho_{\a,\bd}(x,Z;Y,\th)$, so a residual gauge transformation
parameter $\he(x,Z;Y,\th)$ obeys $\d_{\he}\ho_{\a}|_{Z=0}= (\del_\a+ 2i
\o_{\a}{}^{\b} \del_\b+\cdots)\he|_{Z=0}=0$. The \alg\ valued and
therefore $Z$-independent parameter $\he=i_*\e$ clearly obeys this
condition.}
of \eq{hoa}, which are

\bea
\d_{\vare}\o&\se& i^*\d_{i_*\vare}\ho \se d\vare -i^*\sc{\ho}{i_*\vare}\ ,
\la{wgt}\w2
 \d_{\vare}\f&\de& i^*\d_{i_*\vare}\hf\se i^*(i_*\vare\star \hf-\hf\star
\bp i_*\vare)\ ,
\la{figt}
\eea

where $\vare(x;Y,\th)$ is an \alg\ valued parameter. From the closure
of the $\d_{\he}$ transformations on $\ho$ and $\hf$, and using the
property $\sc{i_*\vare_1}{i_*\vare_2}=i_*\sc{\vare_1}{\vare_2}$, it
follows that the residual gauge transformations \eqs{wgt}{figt} close.

Let us now return to the twisted reality condition \eq{twr}. To
illustrate the role of the operator $K$ in \eq{twr}, let us suppose
that we have a solution to \eqs{e1}{e6} such that $\hf^{(0)}=0$. We can
then set $\ho^{(0)}_\a=0$ by means of an \alge\ gauge transformation.
Then \eq{e4} implies that $\del_\a\ho^{(0)}=0$ which is solved by
$\ho^{(0)}_\m=\O^{(0)}_\m(Y,\th)$. Next, we observe from \eq{e1} that a
small fluctuation field $\f^{(1)}$ obeys $\del_\a\hf^{(1)}=0$, which is
solved by $\hf^{(1)}=\f(Y,\th)$. Then \eq{e6} reads

\be
d\f-\O^{(0)}\star\f+\f\star\bp\O^{(0)}\se0\ .
\la{rig2}
\ee

Since $\O^{(0)}$ is flat one might expect that this equation only
allows rigid solutions, but a more careful analysis (see
\sect{sec:phi} in the case when $\O^{(0)}$ is the AdS vacuum)
shows that these equations do not constrain the chiral components
$\f(m,0;\th)$. Moreover, it follows from

\be
\f(y,\yb;\th)\star\kb|_{Z=0}\se \f(y,0;\th)
\ee

that only the chiral components of $\f$ contribute to the linearization
of $\cR$. \eq{z0} thus expresses the linearized spacetime curvatures
$R^{(1)}$ in terms of the chiral components of $\f$ (see
\eqs{weyl1}{cc} in the case when $\O^{(0)}$ is the AdS vacuum)

The crucial role of the $K$-twisted reality condition \eq{twr} can now
be understood as follows. If we impose an ordinary reality condition
without the $K$-twist, namely if we impose the condition
$\e^{\ad\bd}\hR_{\ad\bd}=\e^{\a\b}\hR_{\a\b}$, then
$\hR_{\a\b}=-\ft{i}2 \e_{\a\b}\thf$ where $\thf^\dagger =
\thf$ and the constraint \eq{rig2} now becomes

\be
d\tif-\O^{(0)}\star\tif +\tif\star \O^{(0)} \se 0\ ,
\la{rig}
\ee

where $\tif=i^*\thf$. This constraint does not allow any local dynamics.
This can be seen by observing that \eq{rig} is identical in form to the
higher spin Killing equation for rigid gauge transformations that
preserve the connection $\O^{(0)}$ (the only difference between $\tif$
and $\e$, which is insignificant in the present context, is that they
have opposite $\tau$-parity)
\footnote{
The flatness of $\O^{(0)}(Y,\th)$ implies that $\O^{(0)}=g^{-1}\star
dg$ where $g$ is an \alg\ group element. \eq{rig} is then solved
formally by taking $\tif= g^{-1}\star \tilde{P}\star g$ for an
\alg\ valued polynomial $\tilde{P}$ with constant coefficients. Applying
this formal method to \eq{rig2} gives $\f=g^{-1}\star P\star\bp g$ for
a constant $P$. These results naively would imply that both \eq{rig2}
and \eq{rig} have only rigid solutions, whereas a more careful analysis
shows that there is room for unconstrained chiral components in the
solutions to \eq{rig2} while \eq{rig} indeed only allows the rigid
solutions.}.

We conclude this section by noting that the constraints \eq{master} and
\eq{dhf} are invariant under the spacetime diffeomorphisms which are
incorporated into the \alge\ gauge group such that $\d x^\m=\x^\m$ is
generated by the gauge transformation with parameter $\he(\x)=i_\x\ho$.
The $Z$ space diffeomorphisms, on the other hand, are broken down to
local $SO(3,1)$ rotations by the reality condition \eq{twr}. To
preserve local $SO(3,2)$ invariance one would have to impose an
untwisted reality condition which would not yield dynamics as argued
under \eq{rig}. The untwisted reality condition that leads to \eq{rig}
is equivalent to the vanishing of $(\c^{\una})^{\ua\ub}\hR_{\ua\ub}$
where $(\c^{\una})_{\ua\ub}\ (\una=0,1,2,3,5)$ are the $SO(3,2)$
$\c$-matrices.

\tc


\section{The Anti-de-Sitter Vacuum and Spectrum}\la{sec:vac}


\subsection{Expansion Around AdS Vacuum}

The AdS vacuum solution is given by

\bea
\hf^{(0)}&\se& 0\ ,
\nn\w2
\ho_\a^{(0)}&\se& 0\ ,
\nn\w2
\ho_\m^{(0)}&\se& \O^{(0)}_{\m}\de\ft1{4i}\left[\o_{\m\a\b}^{(0)}
y^{\a}y^{\b}+\bo_{\m\ad\bd}^{(0)} \yb^{\ad}\yb^{\bd} +
2e^{(0)}_{\m\a\ad}y^\a\yb^{\ad}\right]\ ,
\la{vac}
\eea

where $\o^{(0)}=\ft1{4i}(\o^{(0)}_{\a\b}y^\a y^\b+h.c)$ is the
$SO(3,1)$ valued Lorentz connection of AdS spacetime of radius
$\l^{-1}$ and and with vierbein $e^{(0)\;a}= \l^{-1}
(\sb^a)^{\ad\a}e^{(0)}_{\a\ad}$. The $SO(3,2)$ curvature vanishes, that
is $d\O^{(0)}-\O^{(0)}\star \O^{(0}=0$. In $SO(3,1)$ covariant
component notation this means

\bea
d\o^{(0)}_{\a\b}&\se&\o^{(0)}_{\a\c}\wedge\o^{(0)}_{\b}{}^{\c}+
e^{(0)}_{\a\dd}\wedge e^{(0)}_{\b}{}^{\dd}\ ,
\nn\w2
d\bar{\o}^{(0)}_{\ad\bd}&\se&\bar{\o}^{(0)}_{\ad\cd}\wedge
\bar{\o}^{(0)}_{\bd}{}^{\cd}+e^{(0)}_{\d\ad}\wedge e^{(0)\,\d}{}_{\bd}\ ,
\nn\w2
de^{(0)}_{\a\bd}&\se&\o^{(0)}_{\a\c}\wedge e^{(0)\,\c}{}_{\bd}
+\bar{\o}^{(0)}_{\bd\dd}\wedge e^{(0)}_{\a}{}^{\dd}\ .
\la{ads}
\eea

We shall use a notation where

\be
\nab\de e^{(0)\,a}\nab_a
\la{nab}
\ee

is the $SO(3,1)$ covariant derivative with connection $\o^{(0)}$. Its
curvature, i.e. the Riemann tensor, is given by $r^{(0)}_{ab,cd}=
-\l^2(\y_{ac}\y_{bd}-\y_{ad}\y_{bc})$. The AdS vacuum is invariant
under rigid gauge transformations
\footnote{The Killing equation reads $d\vare^{(0}
-\sc{\O^{(0)}}{\vare^{(0)}}=0$ and in the AdS case it consists of
decoupled systems of equations for $\vare^{(0)}(m,n;\th)$ for each
value of $m+n=0,1,2,...$, and these equations admit only rigid
solutions.}
with \alg\ valued parameters $\e^{(0)}$ which can be expressed in terms
of the usual commuting $SO(3,2)$ Killing spinors $\y^r_{\a}$
($r=1,..,4$) obeying

\be
\nab\eta^r_{\a}\se e^{(0)}_{\a\ad}\bar{\y}^{r\;\ad}\ .
\la{iso}
\ee

Defining $\y^r:=\y^r_\a y^\a- \y^r_{\ad}\yb^{\ad}$, the rigid parameter
$\vare^{(0)}$ is then an arbitrary \alg\ valued polynomial in $\y^r$,
$\yb^r$ and $\th^i$ with constant coefficients.

Solving for the $Z$ dependence $\hf$ and $\ho$ from \eqs{e1}{e4}, as
explained in \sect{sec:ga}, in a perturbation expansion around the AdS
vacuum solution \eq{vac} and then substituting these solutions into
\eq{z0} one obtains its linearized form, which in $SO(3,1)$ covariant
notation is given by
\footnote{We use the following symmetrization convention: all $SO(3,1)$
spinor indices indices of the same type, such as $\a_1,\a_2,...\a_m$,
are to be understood to be symmetrized with unit strength.}

\bea
&& R^{(1)}_{\a\b,\mi\c1{2s-2}}(\th\,)\se \f_{\a\b\mi\c1{2s-2}}(\th\,)
\ ,\quad\quad s=1,\ft32,2,...\ ,
\la{weyl1}\w4
&& R^{(1)}_{\ad\bd,\mi{\cd}1{2s-2}}(\th\,)\se
\f_{\ad\bd\mi{\cd}1{2s-2}}(\th\,)\star \C\ ,\qq  s=1,\ft32,2,...\ ,
\la{weyl2}\w4
&& R^{(1)}_{\a\b,\mi\c1k\mi{\cd}{k+1}{2s-2}}(\th)\se 0\ ,\qq
\quad\quad\quad s=\ft32,2,\ft52,...\ ,\quad k=0,..,2s-3
\la{cc}\w4
&& \nab_{\a\ad}\f_{\mi\b1m\mi{\bd}1n}(\th\,)\se
i\,\f_{\a\mi\b1m\ad\mi{\bd}1n}(\th\,)-i\,mn\,\e_{\a\b_1}\,\e_{\ad\bd_1}\,
\f_{\mi\b2m\mi{\bd}2n}(\th\,)
\nn\w2
&&\qquad \qquad\qquad\qquad\qquad\qquad\qquad
\qquad\qq \ \  m,n=0,1,2,...\la{fc}
\eea

where $R^{(1)}=d\o-\{\O^{(0)},\o\}_{\star}$ is the linearization of the
spacetime curvature $R$ given in \eq{rd} and it reads

\bea
&& R^{(1)}_{\a_1\a_2,\mi\b1m\mi{\bd}1n}(\th\,)
\de 2\, \nab\o_{\a_1\a_2,\mi\b1m\mi{\bd}1n}(\th\,)
\nn\w2
&& -m\,\e_{\a_1\b_1}\,
      \o_{\a_2}{}^{\cd}{}_{,\mi\b2m\cd\mi{\bd}1n}(\th\,)
   -n\,\o_{\a_1\bd_1,\a_2\mi\b1m\mi{\bd}2n}(\th\,)\ ,
\la{b1}
\eea

which obeys the linearized Bianchi identity
$dR^{(1)}-\sc{\O^{(0)}}{\o}=0$, that is

\bea
&& \nab_{\ad}{}^{\c}R^{(1)}_{\c\a,\mi\b1m\mi{\bd}1n}-
   \nab_{\a}{}^{\cd}R^{(1)}_{\cd\ad,\mi\b1m\mi{\bd}1n}
\nn\w2
&& - m\left[R^{(1)}_{\a\b_1,\mi\b2m\ad\mi{\bd}1n}-\e_{\a\b_1}
R^{(1)}_{\ad}{}^{\cd}{}_{,\mi\b2m\cd\mi{\bd}1n}\right]
\nn\w2
&& +n\left[R^{(1)}_{\ad\bd_1,\a\mi\b1m\mi{\bd}2n}-\e_{\ad\bd_1}
R^{(1)}_{\a}{}^{\c}{}_{,\c\mi\b1m\mi{\bd}2n}\right]\se0\ .
\la{b2}
\eea

We have converted the curved spacetime indices of the forms into flat
indices using the AdS vierbein $e^{(0)\;a}_{\m}$ and set

\bea
\o_{\a\bd}&\de&(\s^{a})_{\a\bd}\,\o_{a} \ ,\qq
R^{(1)}_{\a\b}\de\ft12(\s^{ab})_{\a\b}R^{(1)}_{ab} \ ,
\nn\w2
\nab_{\a\ad}&\de&(\s^a)_{\a\ad}\nab_{a}\ ,\qq
\nab\o_{\a\b}\de \ft12(\s^{ab})_{\a\b}\,\nab_{a}\o_{b}\ .
\la{defns}
\eea

The linearized field equations \eqs{weyl1}{b2} are invariant under the
linearization of the \alg\ valued gauge transformations \eqs{wgt}{figt},
which assume the following form in $SO(3,1)$ covariant component language:

\bea
\d\o_{\a\ad,\mi\b1m\mi{\bd}1n}&\se& \nab_{\a\ad}\vare_{\mi\b1m\mi{\bd}1n}
+m \,\e_{\a\b_1}\vare_{\a\mi\b2m\ad\mi{\bd}1n}
\nn\w2
&&+n\,\e_{\ad\bd_1}\vare_{\mi\b1m\mi{\bd}2n}\ ,
\la{g}\w2
\d\f&\se& 0\ .
\la{gf}
\eea

\vspace{1cm}

\tf
%


\subsection{The Spin $s\leq 1$ Linearized Field Equations}\la{sec:phi}


We begin by analyzing \eq{fc} from which we find that the non-chiral
spin $s+l$ component $\f(2s+l,l;\th)\ (l=1,2,...)$ is expressed in
terms of its chiral component $\f(2s,0;\th)$ as follows

\be
\f_{\mi\a1{2s+l}\mi{\ad}1l}(\th)\se
\nab_{\a_1\ad_1}\nab_{\a_2\ad_2}\cdots \nab_{\a_l\ad_l}
\f_{\mi\a1{2s}}(\th)\ .
\la{fsolv}
\ee

The chiral components of $\f$ are identified with the chiral curvature
components as in \eq{weyl1}. These chiral curvature components are the
Weyl tensors and they are the only spacetime curvature components that
are non-vanishing on-shell, as follows from \eqs{weyl1}{weyl2}.
Inserting the expansions \eq{wexp} and \eqs{fimn}{mm} into
\eq{weyl1} and \eq{fsolv} one verifies that the $SO(8)$ content of
both sides of these equations indeed match. Thus, the spin $s\geq1$
components of the master field $\f$ are auxiliary in the sense that
they can be solved for algebraically in terms of the gauge fields and
their derivatives. The spin $s\leq \ft12$ components, on the other
hand, are independent since they cannot be eliminated algebraically.

In obtaining \eq{fsolv} one uses only the fully symmetrized components
of \eq{fc}. The remaining components are $\e$-traces. By combining
\eqs{weyl1}{weyl2} with the Bianchi identity
\eq{b2} for $s\geq\ft32$ one can show by induction that the $\e$-traces of
\eq{fc} for $m+n\geq3$ are {\it identically} satisfied provided that its
$\e$-traces for $m+n\leq 2$ are satisfied. These constitute the
linearized equations of motion for the spin $s\leq1$ fields and the
Bianchi identity for the spin $1$ fields, which we summarize as
follows:

\begin{itemize}

\item[a)] The complex scalar $\f(0,0)$ and the $35_++35_-$ scalars
$\f^{ijkl}(0,0)$ are dynamical fields which constitute $\f(\th)\equiv
\f(0,0;\th)$ (see \eq{mm}) which obey the linearized field
equation $(\nab_{\a}{}^{\ad}\nab_{\b\ad}+2\e_{\a\b})\f(\th)=0$, that is

\be
(\nab^2+2)\f(\th)\se0\ .
\la{scalar}
\ee

\item[b)] The complex Weyl fermions $\f^{ijk}_\a$ and
$\f^{\mi{i}17}_\a$ are dynamical fermionic fields which make up
$\f_{\a}(\th)$ which obey the linearized field equation
$\nab_{\ad}{}^\a\f_\a(\th)=0$, that is

\be
(\sb^a)^{\ad\a}\nab_a \f_{\a}(\th)\se 0\ .
\la{dirac}
\ee

\item[c)] The spin $1$ field strength $R^{(1)}_{\a\b}(\th)=\f_{\a\b}(\th)$,
obeys $\nab_{\ad}{}^{\b}R^{(1)}_{\b\c}(\th)=0$, which amounts to

\be
\nab^aR^{(1)}_{ab}(\th)\se 0 \ ,\qquad
\e^{abcd}\del_b R^{(1)}_{cd}(\th)\se0\ .
\la{vector}
\ee

It follows that $R^{(1)}(\th)=d\o(\th)$, and in the Lorentz gauge
$\nab^a\o_a(\th)=0$ the resulting second order vector equation reads

\be
(\nab^2+3)\,\o(\th)\se0\ .
\ee

\end{itemize}

Hence, taking into account the rank of the Dirac operator in \eq{dirac}
and the decoupling of the longitudinal mode in the spin $1$ sector, one
finds that there are two real physical component functions for each
spin and $SO(8)$ label. These have mode function expansions with modes
making up the massless representations of the AdS algebra, that is the
representations $D(E_0=s+1,s)$ for $s=\ft12,1$ and $D(1,0)$ or $D(2,0)$
for $s=0$. We use a notation where $D(E_0,s)$ denotes a representation
of $SO(3,2)$ in which $E_0$ labels the minimum energy eigenvalue of the
energy operator $M_{05}$ and $s$ is the maximum eigenvalue of the spin
operator $M_{12}$ in the lowest energy sector.


\subsection{Auxiliary and Dynamical Gauge Fields and
Symmetries}\la{sec:aux}


We next turn to solving the vanishing curvatures in \eqs{weyl1}{cc}
modulo the relationships among them due to the Bianchi identity
\eq{b2}. The end result, which is illustrated in Fig. (\ref{wfig}), is
that the {\it auxiliary gauge fields} $\o(m,n;\th)$ for $|m-n|\geq 2$
are given in terms of the {\it dynamical gauge fields} $\o(m,n;\th)$
for $|m-n|\leq1$ as follows:

\bea
&&\o_{\a\ad,\mi\b1m\mi{\bd}1n}(\th)\se \ft2{m+1}\nab\o_{\bd_1\bd_2,\a\mi\b1m
\ad\mi{\bd}3n}(\th)
\nn\w2
&&+\e_{\ad\bd_1}\ft{2n}{n+1}\left[\ft{n-1}{m+n+2}\nab\o_{\bd_2}
{}^{\cd}{}_{,\a\mi\b1m\cd\mi{\bd}3n}(\th)\right.
\nn\w2
&&+\ft{n+1}{m+n+2}  \nab\o_{(\a}{}^{\c}{}_{,\mi\b1m)\c\mi{\bd}2n}(\th)
\nn\w2
&&\left.-\ft{m}{(m+1)(m+2)}\e_{\a\b_1}\nab\o^{\c\d}{}_{,\c\d\mi\b2m
\mi{\bd}2n}(\th)\right]
\nn\w2
&&+m\,\e_{\a\b_1}\x_{\mi\b2m\ad\mi{\bd}1n}(\th)\ ,\qquad n-2\geq m\geq0\ .
\la{auxsolv}
\eea

The components $\x(m-1,n+1,\th)$ ($m>0$)  are not determined by the
constraint. Thus the auxiliary gauge fields $\o(m,n;\th)$ are given by
$|m-n|-a$ derivatives of the dynamical gauge fields where $a=1$ for
bosonic fields and $a=2$ for fermionic fields.

We refer to the dynamical gauge fields $\o(s-1,s-1;\th)$ ($s=2,3,...$)
as the generalized vierbeins and the dynamical gauge fields
$\o(s-\ft32,s-\ft12;\th)$ ($s=\ft32,\ft52,...$) as the generalized
gravitini. The auxiliary gauge fields $\o(s,s-2;\th)$ ($s=2,3,...$) are
referred to as the generalized Lorentz connections.

The local gauge symmetries \eq{g} are divided into auxiliary gauge
symmetries with parameters $\vare(m,n;\th)$ for $|m-n|\geq 3$ and
dynamical gauge symmetries with parameters $\vare(m,n;\th)$ for
$|m-n|\leq2$. The dynamical gauge symmetries in the bosonic spin $s$
sector are the generalized reparametrizations with parameters
$\vare(s-1,s-1;\th)$ and the generalized local Lorentz transformations
with parameters $\vare(s-2,s;\th)$, and in the fermionic spin $s$
sector they are the generalized local supersymmetries with parameters
$\vare(s-\ft32,s-\ft12;\th)$ and the fermionic, Lorentz-like
transformation with parameters $\vare(s-\ft52,s+\ft12;\th)$. From
\eq{g} it follows that

\be
\d_{\vare_{aux}}\o_{dyn}\se0\ ,
\la{auxg}
\ee

where $\vare_{aux}$ is a parameter of an auxiliary gauge transformation
and $\o_{dyn}:=\sum_{|m-n|\leq1}\o(m,n;\th)$. The auxiliary gauge
transformation with parameter $\vare(m-1,n+1;\th)$ induces a shift in
the undetermined component $\x(m-1,n+1,\th)$ in the auxiliary gauge
field $\o(m,n;\th)$ given by \eq{auxsolv}. This Stueckelberg type gauge
symmetry can be completely fixed by imposing the gauge condition

\be
\x(m,n;\th)\se0\ .
\ee

The definition \eq{auxg} also implies that the dynamical gauge
symmetries form a subalgebra of \alg\ which closes on the dynamical
fields, since if
$[\vare^1_{dyn},\vare^2_{dyn}]_{\star}=\vare_{dyn}+\vare_{aux}$ then
from \eq{auxg}

\be
[\d_{\vare^{1}_{dyn}},\d_{\vare^2_{dyn}}]\o_{dyn}\se
\left(\d_{\vare_{dyn}} +\d_{\vare_{aux}}\right)\o_{dyn}\se
\d_{\vare_{dyn}}\o_{dyn}\ .
\ee


\subsection{The Spin $s> 1$ Linearized Field Equations}\la{sec:s>1}


The components of the curvature constraints \eqs{weyl1}{cc} that are
not used up in solving for the auxiliary gauge fields are the field
equations for the spin $s>1$ dynamical gauge fields. They read

\bea
&&e^{(0)}_{\a_1}{}^{\bd}\wedge
R^{(1)}_{\mi\a2m\bd\mi{\ad}1n}(\th)\se0\ ,
\nn\w2
&&\ba{ll}
m=n&\mbox{for spin}\ s=m+1=2,3,...\w2 m=n+1&\mbox{for spin}\
s=m+\ft12=\ft32,\ft52,...
\ea
\la{eom}
\eea
\w{1}

In the bosonic sector these are first order equations for the
generalized Lorentz connections, which by virtue of \eq{auxsolv} turn
into second order field equations for the generalized vierbeins. In the
fermionic sector \eq{eom} are first order Dirac like field equations
for the generalized gravitini.

By making use of the dynamical gauge symmetries one can fix the
generalized Lorentz type gauges

\bea
&&\nab^{\a\ad}\o_{\a\ad,\mi\b1{s-1}\mi{\bd}1{s-1}}(\th)\se0\ ,
\nn\w2
&&\o_{\ad}{}^{\b}{}_{,\b\mi\b1{s-2}\mi{\bd}1{s-1}}(\th)\se 0\ ,\qquad
s=2,3,...
\la{lg}
\eea

in the bosonic sector, and

\bea
\nab^{\a\ad}\o_{\a\ad,\mi\b1{s-3/2}\mi{\bd}1{s-1/2}}(\th)&\se&0\ ,
\qquad s=\ft32,\ft52,...
\nn\w2
\o_{\a}{}^{\ad}{}_{,\mi\b1{s-3/2}\ad\mi{\bd}1{s-3/2}}(\th)&\se&0\ ,
\qquad s=\ft32,\ft52,...
\nn\w2
\o^{\a}{}_{\bd_1,\a\mi\b1{s-5/2}\mi{\bd}2{s+1/2}}(\th)&\se&0\ ,
\qquad s=\ft52,\ft72,...
\la{fgf}
\eea

in the fermionic sector. These conditions eliminate all $SO(3,1)$
irreducible components in the generalized vierbeins and gravitini
except the spin $s$ components $\o_{\a_1\ad_1,\mi\a2s\mi{\ad}2s}(\th)$
and $\o_{\a_1\ad_1,\mi\a2{s-1/2}\mi{\ad}2{s+1/2}}(\th)$, respectively,
for which one obtains the field equations

\be
\left[\nab^2+3-(s-1)^{2}\right]\o_{\a_1\ad_1,\mi\a2s\mi{\ad}2s}(\th)\se0\ ,
\la{gvbe}
\ee

for spin $s=2,3,...$, and

\be
\nab_{\a}{}^{\cd}\o_{\b_1\cd,\mi\b2{s-1/2}\mi{\bd}1{s-1/2}}(\th)-
(s-\ft12)\o_{\a\bd_1,\mi\b1{s-1/2}\mi{\bd}2{s-1/2}}(\th)\se 0\ ,
\la{de}
\ee

for spin $s=\ft32,\ft52,...$ These equations describe massless
representations $D(E_0=s+1,s)$ of the AdS algebra, as explained at the
end of \sect{sec:phi}.

Any one of these representations are characterized by two real
transverse functions. To see this we first note that in the bosonic
sector the gauge condition \eq{lg} has residual gauge transformations
given by generalized reparametrizations obeying

\be
[\nab^{2}+1-s^2]\vare_{\mi\b1{s-1}\mi{\bd}1{s-1}}(\th)\se0\ ,\qquad
\nab^{\a\ad}\vare_{\a\mi\b1{s-2}\ad\mi{\bd}1{s-2}}(\th)\se0\ .
\la{resb}
\ee

These gauge transformations obey the gauge fixed equations of motion
\eq{gvbe} and therefore decouple the longitudinal states from the
physical spectrum. Hence the number of real transverse bosonic spin $s$
functions is given by the number of components of the spin $s$ irrep
($(s+1)^2$), minus the number of gauge conditions in \eq{lg} that are
linear in derivatives ($s^2$), minus the number of residual gauge
symmetries, which is equal to the number of degrees of freedom in
$\vare(s-1,s-1)$ ($s^2$) minus the number of constraints in \eq{resb}
that are linear in derivatives ($(s-1)^2$), which indeed sum up to
$(s+1)^2-s^2-[s^2-(s-1)^2]=2$. Similarly, in the fermionic sector one
has residual generalized local supersymmetries obeying

\be
\nab_{\a}{}^{\cd}\vare_{\mi\b1{s-3/2}\cd\mi{\bd}1{s-3/2}}(\th)-
(s+\ft12)\,\vare_{\a\mi\b1{s-3/2}\mi{\bd}1{s-3/2}}(\th)\se0\ .
\la{resf}
\ee

Hence, taking into the account the fact that the Dirac operator has
half the maximal rank, one finds that for any occurring $SO(8)$
representation there are indeed
$(s+\ft12)(s+\ft32)-(s-\ft12)(s+\ft12)-[(s-\ft12)(s+\ft12)-
(s-\ft32)(s-\ft12)]= 2$ real transverse fermionic spin $s$ functions.

Using rigid supersymmetries with parameter
$\vare^{(0)}=\ft1{2i}\vare_{\a i}y^{\a}\th^i+h.c$ constructed from the
Killing spinors in \eq{iso} by setting $\vare_{\a i}=
\vare_{ri}\y^r_{\a}$, we find that the physical spectrum obtained in
\eqs{scalar}{vector} and \eqs{gvbe}{de} can be arranged into the tower
of supermultiplets given in Table \ref{table}. The lowest level,
$\ell=0$, is the $CPT$ self-conjugate short supergravity multiplet,
while the higher levels each consists of two short irreducible
multiplets that are transformed into each other under $CPT$. Let us
discuss the structure of the level $\ell=0$ multiplet in more detail.

\tw
%


\subsection{Embedding of the $N=8$ Gauged Supergravity}\la{sec:com}


From \eqs{scalar}{vector} and \eq{eom} it follows that the linearized
equations of motion of the level $\ell=0$ multiplet are given by

\bea
\mbox{spin $s=2$} &:&\qquad r^{(1)}_{ab}\se 0\ ,
\nn\w2
\mbox{spin $s=\ft32$} &:&\qquad(\s^{abc})_{\a}{}^{\ad}\left(
\nab_{b}\,\o_{c,\ad}^{\,i}-\ft12\l\,(\s_b)_{\ad}{}^{\b}\o_{c,\b}^{\,i}\right)
\se0\ ,
\nn\w2
\mbox{spin $s=1$} &:&\qquad \nab^{a}\del_{[a}\o_{b\,]}^{\,ij}\se0\ ,
\nn\w2
\mbox{spin $s=\ft12$} &:&\qquad
\nab_{\a}{}^{\ad}\f_{\ad}^{ijk}\se0\ ,
\nn\w2 \mbox{spin $s=0$} &:&\qquad
\left(\nab^2+2\l^2\right)\f^{ijk\,l}\se0\ ,
\la{sgm}
\eea

where $\l$ is the inverse AdS radius and $r^{(1)}_{ab}$ is the
linearization of the Ricci tensor $r_{ab}=
e_{a}{}^{\m}e_{c}{}^{\n}r_{\m\n,bc}$ where $r_{\m\n,ab}$ is the
$SO(3,1)$ valued Riemann curvature. These equations follow from varying
the quadratic part of the gauged $N=8$ supergravity action \cite{dn1}

\bea
&&e^{-1}{\cal L}_2\se {1\over 2} R -{\k^{2}\over 8g^{2}}
F_{\m\n,IJ}F^{\m\n,IJ} -{1\over 96}
\del_{\m}\phi_{ijkl}\del^{\m}\phi^{ijkl} +{g^{2}\over 24\k^{2}}\,
\phi_{ijkl}\phi^{ijkl} +{6g^2\over\k^2}
\nn\w2
&&\ +\left({i\over 2}
\bar{\psi}_{L\m}^{i}\c^{\m\n\r}\nab_{\n}\psi_{L\r i} +
{ig\over\sqrt{2}\k}\bar{\psi}^{i}_{R\m}\c^{\m\n}\psi_{L\n,i} -{1\over
2} {\bar\chi}^{ijk}_L\c^{\m}\nab_{\m}\chi_{Lijk}\ +h.c.\right) \ ,
\la{lagr}
\eea

where $i,j,..=1,...,8$ are $SU(8)$ indices and $I,J,..=1,...,8$ are
$SO(8)$ indices, $\phi_{ijkl}$ are the $35+35$ scalars obeying the
$SU(8)$ reality condition \eq{mm}
and the fermions are Weyl. The complex conjugation changes chirality,
and consequently both chiralities occur for the gravitini as well as
the spin $\ft12$ fields. Thus, the theory is vector like. In writing
\eq{lagr} we have assigned (energy) dimension $\ft12$ to all the
fermions, dimension $0$ to the scalars and dimension $1$ to the vector
fields. We find that \eq{sgm} is in perfect agreement with \eq{lagr}
provided that we make the identifications

\bea
\o_\m{}^{\a\ad} \ \ &\rightarrow& \ \ \l\,e_\m{}^a\ ,\qquad\
\o_\m^{ij} \ \ \rightarrow \ \ g\,A_\m^{IJ}\ , \nn\w2 \o_{\m\a}^i \ \
&\rightarrow& \ \ \l^{\ft12}\ ,\psi_{L\m}^i\ ,\qquad
\l^{-ft12}\,\f_\a^{ijk}
\
\
\rightarrow
\
\
\chi_L^{ijk}\ ,
\eea

and identify the following important relation between the Newton's
constant $\k$, the $SO(8)$ gauge coupling $g$ and the inverse AdS
radius:

\be
{g^2\over \k^2} \se \ft12\l^2\ .
\la{gk}
\ee

The free parameters of the higher spin theory are therefore the gauge
coupling $g$ and the inverse AdS radius $\l$. The gauge coupling is
introduced into the full set of higher spin equations \eqs{e1}{e6} by
replacing $\ho\rightarrow g\, \ho$. These equations are consistent with
the assignment of dimension $0$ to the master fields $\ho$ and $\hf$
provided that one makes use of the dimensionful parameter $\l$ to
rescale the component fields. Consequently the interaction terms will
have negative powers of $\l$\cite{fv1,fv2} and therefore the flat limit
in which $\l\rightarrow 0$ is singular.


\section{On the Uniqueness of the Constraints}\la{sec:uni}


A natural question is whether there are curvature constraints other
than \eq{master} that realize the
\alg\ dynamics irreducibly. However, the uniqueness of the $N=8$, $D=4$
gauged supergravity suggests that there should not be much room for
interaction ambiguity. Nonetheless, one might relax \eq{master}. A
relaxation that preserves irreducibility amounts to introducing more
auxiliary fields over and above those already present in $\o_\m(Y,\th)$
and $\f(Y,\th)$. A relaxation that does not maintain irreducibility
could provide room in the curvature for propagating \alg\ multiplets
other than the gauge multiplet. These two types of generalizations
could be useful in finding an action or matter couplings.

To relax the condition on $\hR_{\m M}$ while preserving the local
symmetries one must first convert the curved spacetime indices on
$\hR_{\n M}$ into flat indices using a preferred invertible gauge field
such as the vierbein and then find the relaxed \alge\ valued
constraint. Insisting on the Yang-Mills like form of the \alge\ gauge
transformations this does not seem possible.

Turning to the remaining constraints, setting $\hR_{\a\bd}=
(\s^a)_{\a\bd}\hf_a ,\ (a=0,1,2,3)$ modifies \eq{dhf} such that
$\hD_\a\hf$ will be proportional to
$(\s^a)_{\a}{}^{\bd}(\hD_{\bd}\hf_a)\star \kb$. There are further
integrability conditions involving $Z$-dependence and the (untwisted)
condition $\hD_\m\hf^a=0$. The full consequences of introducing the
field $\hf^a$ is not clear to us at present.

Finally we consider modifying the constraint \eq{master} as

\be
\hR\se \ft{i}4\left[dz^\a\wedge dz_\a \cV(\hf\star \k\C) +
d\zb^{\ad}\wedge d\zb_{\ad} \bcV(\hf\star \k^{\dagger})\right]\ ,
\la{gm}
\ee

where $\cV$ is an arbitrary $\star$ function and $\hf$ obeys \eq{hfi}.
The Bianchi identities $D_{\ad}\cV(\hf\star\k\C)=0$ and
$D_\a\bcV(\hf\star\kb)=0$ are integrable conditions which imply the
condition \eq{dhf}. The constraint \eq{gm} implies the non-linear
reality condition

\be
\e^{\ad\bd}\hR_{\ad\bd}\se\bcV(\cV^{-1}(\e^{\a\b}\hR_{\a\b})\star K)\ .
\la{vv}
\ee

The constraint \eq{gm} evidently describes a system with the same
degrees of freedom as the system described by \eq{master}. To analyze
further the consequences of \eq{gm} let us consider the field
redefinition

\be
\hf'\se \cF(\hf\star \k\C)\star \k\C\ ,
\la{fr}
\ee

where $\cF$ is a $\star$ function. Defining $\cF_{\pm}(-X):=\pm\cF(X)$
and observing that $\cF_{-}(\hf\star\k\C)=
\cF_{-}(\hf\star\kb)\star\k\C$ and $\cF_{+}(\hf\star\k\C)=
\cF_{+}(\hf\star\kb)$, we find from
\eq{dhf} that

\be
D_{\ad}\hf'\se 0\ ,\qquad D_\a\hf'\se\cF_{+}\star D_\a K\ .
\la{dh}
\ee

Moreover, $\hf'$ inherits a reality condition from $\hf$ which in
general is non-linear. For real $\cF$ it simplifies to:

\be
\hf'^{\dagger}\se \p(\hf'_-+\hf'_+\star K)\star\C\qquad
\mbox{for}\quad\cF=\bar{\cF}\ ,
\la{d}
\ee

where $\hf'_{\pm}:=\cF_{\pm}(\hf\star\k\C)\star \k\C$. Thus, there is a
set of apparently inequivalent (see below) constraints given by \eq{gm}
where $\cV$ is a complex $\star$ function defined modulo the
equivalence relation

\be
\cV\;\;\sim\;\;\cV'\qquad \mbox{if}\qquad \cV=\cV'\circ \cF\quad
\mbox{for}\quad
\cF_+=0\ ,\quad \cF=\bar{\cF}\ .
\ee

In other words the odd real part of $\cV$ in \eq{gm} can be redefined
away. Note that if one attempts to set $\cV(X)=X$ by a field
redefinition, then the 'interactions' from its even or imaginary parts
resurface in the reality condition on $\hf$ and the right hand side of
the expression for $D_{\a}\hf$. However, other types of field
redefinitions reduce the amount of freedom in $\cV$. This can be seen
even at the linearized level in a simple example as follows.

Starting from $\cV(X)=aX+\cdots$, where $a=|a|e^{i\th}$, a real field
redefinition $\hf\rightarrow |a|^{-1}\hf$ leads to $\cV(X)=e^{i\th}X$.
However, at the linearized level the phase factor only enters the
curvature constraints \eqs{weyl1}{weyl2} and can therefore be
eliminated by redefining the initial condition $\f$. In the notation
$\f\equiv\f_++\f_-+\f_0$, where $\f_{\pm}(m,n;\th)=0$ for $\pm(m-n)>0$
and $\f_0(m,n;\th)=0$ for $m\neq n$, the phase factor can be absorbed
into the complex fields in $\f_{\pm}$ by redefining
$\f_{\pm}\rightarrow e^{\mp i\th}\f_{\pm}$. Notice that unlike
\eq{fr} these redefinitions treat $\f_{\pm}$ and $\f_0$ differently.


\section{Discussion}\la{sec:dis}


Despite the fact that the higher spin AdS supergravity theory based on
\alg\ has no flat space limit, such a limit may be achieved by
spontaneous breaking of the higher spin symmetries, followed by sending
the AdS radius to infinity in a suitable way.

Another consequence of spontaneous symmetry breaking is that the higher
spin theory may be truncated to the gauged or ungauged $N=8,D=4$
supergravity by sending the symmetry breaking parameter to infinity.
This truncation is similar to the decoupling of gravity from matter by
setting Newton's constant $\k$ equal to zero, in which case one does
not necessarily require that the coupled field equations allow
consistent truncation to the matter sector for finite $\k$.

Spontaneous symmetry breaking may require scalars with a suitable
potential and couplings to the higher spin gauge fields. In particular,
symmetry breaking that allows truncation down to $N=8,D=4$ supergravity
might be achieved by using the $2$ scalars at the first level. However,
a constant value for this field does not appear to solve the field
equations without involving higher spin fields. Therefore a more
plausible scenario is to couple the theory to massive higher spin
multiplets containing fields which can trigger the symmetry breaking.

The $N=8$ supersingleton propagating at the boundary of AdS spacetime,
assumed to arise from the $AdS_4\times S^7$ compactification of the
supermembrane, serves as a spectrum generating representation for the
massless higher spin theory propagating in the bulk of the AdS
spacetime. As mentioned in the introduction, it has been conjectured
that the singletons play a fundamental role in the description of the
higher spin supergravity theory in the bulk. Since the massless spectra
of the bulk and the boundary theories agree, the essential test of the
bulk/boundary duality is whether it is possible to represent the \alg\
symmetry algebra as charges of anomaly free currents in the $N=8$
supersingleton theory. We expect that the nonlinearities of the bulk
theory would then be reproduced by the interactions between composite
singleton states. The higher spin gauge coupling $g$ can be introduced
in the boundary theory by rescaling the composite operators
corresponding to the bulk gauge fields. The boundary theory depends on
$\l$ as well, and the relation \eq{gk} should be predicted by the
boundary interactions. The boundary singleton theory also involves
massive composite states forming infinite dimensional higher spin
multiplets. Therefore the proposed bulk/boundary duality also yields
higher spin bulk interactions of both massless and massive sectors
which could trigger the spontaneous breaking of \alg. While this
program by and large remains to be realized, the construction of higher
spin currents in the boundary of $AdS_5$ has been recently investigated
\cite{anselmi,fz}.

Another interesting issue is whether it is possible to accommodate the
auxiliary spinor variable $Z_{\ua}$ in the boundary theory. The higher
spin curvature constraint might conceivably arise from the $\k$-like
symmetry a by a supermembrane worldvolume embedded into a target space
extended by $Z$-like dimensions.

The generalization of the interacting higher spin field theory to
higher dimensions is another open problem. The intimate relations
between $M2$ and $M5$-branes makes it particularly interesting to study
higher spin theory in $AdS_7$. The $(2,0)$ tensor multiplet $D=6$ is
the doubleton representation of the $AdS_7$ supergroup and its
symmetric tensor product yields an infinite dimensional multiplet of
massless higher spins in $AdS_7$\cite{g6}. The superdoubleton theory
living at the boundary of $AdS_7$ would now arise from the $AdS_7\times
S^4$ compactification of the $M5$-brane, and we expect that it
generates the dynamics of the higher spin extension of the maximal
gauged supergravity in $D=7$.

It is tantalizing to consider the possibility of a massless higher spin
field theory directly in $D=11$. Even though a standard AdS supergroup
does not exist in $D=11$, a generalized version of such a group and its
singleton like representations have been constructed \cite{g5}. Whether
an action, or equations of motion, for these representations can be
written down in the $D=10$ boundary of $AdS_{11}$ remains to be seen.
In this context, it is interesting to note that the representations of
$SO(9)$ as the little group classifying the massless degrees of freedom
of $D=11$ supergravity has been studied \cite{ramond} with interesting
results that suggest the existence of higher spin massless fields in
$D=11$.

\bigskip\bigskip
\noindent{\large \bf Acknowledments}

\medskip

We wish to thank M. Cederwall, B. Nilsson, M. Duff and M. Vasiliev for
useful discussions and Chalmers University of Technology for their kind
hospitality. This research has been supported in part by NSF Grant
PHY-9722090.

\vfill\eject

\section*{References}


\ed